\documentclass[a4paper,11pt]{article}
\usepackage{amsmath}
\usepackage{graphicx}
\usepackage{amssymb}

\usepackage{cite}

\usepackage[hidelinks]{hyperref}

\hypersetup{
  colorlinks = true,     % Colours links instead of ugly boxes
  urlcolor = blue,       % Colour for external hyperlinks
  linkcolor = blue,      % Colour of internal links
  citecolor = red         % Colour of citations
}

\renewcommand{\vec}[1]{\mathbf{#1}}
\renewcommand{\S}{\mathcal{S}}

\newcommand{\sqmod}[1]{\left|#1\right|^{2}}
\newcommand{\expt}[1]{\left<#1\right>}

\newcommand{\rmi}{\mathrm{i}}
\newcommand{\rme}{\mathrm{e}}
\newcommand{\E}{\epsilon}
\newcommand{\f}{\phi}
\newcommand{\R}{\mathcal{R}}
\newcommand{\Env}{\mathcal{E}}
\newcommand{\w}{\omega}

\newcommand{\bra}[1]{\left<#1\right|}
\newcommand{\ket}[1]{\left|#1\right>}

\newcommand{\iprod}[2]{\left<#1|#2\right>}
\newcommand{\iprodM}[3]{\left<#1\right|#2\left|#3\right>}

\newcommand{\oprod}[2]{\left|#1\right>\left<#2\right|}

\newcommand{\sinc}{\mathrm{sinc}}

\begin{document}

\title{The Concept of Entropic Time:\\ \large{A Preliminary Discussion}}

\author{M.P. Vaughan\footnote{\emph{Corresponding author}: martin.vaughan@physics.org}}

\date{}

\maketitle

\begin{abstract}
The deep connection between entropy and information is discussed in terms of both classical and quantum physics. The mechanism of information transfer between systems via entanglement is explored in the context of decoherence theory. The concept of \emph{entropic time} is then introduced on the basis of information acquisition, which is argued to be effectively irreversible and  consistent with both the Second Law of Thermodynamics and our psychological perception of time. This is distinguished from the notion of \emph{parametric time}, which serves as the temporal parameter for the unitary evolution of a physical state in non-relativistic quantum mechanics. The interpretation of these ideas in terms of both subjective and objective collapse models is also discussed. It is shown that energy is conserved under subjective collapse schemes whereas, in general, under objective collapse it is not. This is consistent with the fact that the latter is inherently non-unitary and that energy conservation arises out of time symmetry in the first place.
\end{abstract}

\tableofcontents

%*******************************************************************************************
\section{Introduction}
%*******************************************************************************************
Quantum mechanics is a unitary theory. A physical system is modelled as a state $\ket{\Psi}$ in a Hilbert space $\mathcal{H}$ and evolves according to a unitary operator $U$ that maps the system at a time $t_{0}$, say, to another at $t_{1}$. The unitary nature of this evolution means that, by operating on the system with the complex transpose $U^{\dagger}$, we can map it back again to its original state. In other words, quantum mechanics is inherently \emph{reversible}.

This reversibility appears fundamentally at odds with the most well-established of all physical laws: the Second Law of Thermodynamics. The essence of this law is that most, if not all, physical processes are \emph{irreversible}, giving a definitive arrow to the direction of time. This direction is associated with an increase in (perhaps the most profound of all physical quantities) \emph{entropy}. The profundity of this concept lies in the fact that it cannot be easily explained in the context of any unitary physical theory, including classical physics. Outside of thermodynamics, the laws of classical mechanics, electromagnetism and quantum physics are time symmetric - they apply equally well if we take time to be evolving in the positive or negative direction. 

In this essay, we discuss and explore the deep connection between the concepts of entropy and information in both classical and quantum physics. On the basis of this, we will introduce a notion of evolution based on the accrual of information, which we refer to as \emph{entropic time}. This is to be distinguished from the usual notion of time as the parameter of the unitary evolution of a system. This latter notion we may refer to as \emph{parametric time}, although in the text we will usually just refer to it as `time'.  In the course of this essay, we come to argue that, whilst we may ordinarily think of the temporal evolution of things in terms of parametric time, this is \emph{not} what constitutes our subjection perception of time. This, we shall argue, must necessarily be based on the phenomenon of information transfer.

The term `entropic time' has previously been coined by Caticha\cite{caticha2011entropic_A} as notion of time ordering emerging naturally out of a theory of entropic dynamics. In Caticha's approach, an attempt is made to derive quantum theory via a scheme of entropic inference based on the conditional probability distributions for particle positions. In our approach, we assume standard quantum theory and develop a concept of entropic time in terms of the information acquired on projective measurement. In both cases, `entropic time' is deeply related to informational content and features a well-defined directionality.

To lay the groundwork for this idea, we begin in Section~\ref{sec:thermodynamic} with a brief overview of thermodynamic entropy as originally introduced by Clausius\cite{clausius1865verschiedene}  before presenting the concept in the statistical form due to Boltzmann\cite{boltzmann1866mechanische} and Gibbs\cite{jaynes1965gibbs}. What appears to be missing from either of these descriptions is how the entropy of the system increases as an explicit function of time on the basis of unitary physics. Indeed, in terms of the Gibbs-Boltzmann formulation, the entropy of a system is given in terms of the probabilities we assign to its microscopic configurations, which we call `micro-states'. However, since the number of micro-states of a system evolving deterministically from an initial micro-state is by necessity unity, the probability of that micro-state will be also be unity and the entropy should never change!

The connection between entropy and the information we have about a system seems to have been first drawn by Maxwell in his famous `demon' thought experiment\cite{maxwell1891theory}, discussed in Section~\ref{sec:information}.  This seems to suggest an epistemological rather than ontological basis for entropy. Such a notion is even more tantalisingly suggested by the mathematical equivalence of the Gibbs entropy and the measure of information introduced by Shannon in communications theory\cite{shannon1948mathematical,shannon2001mathematical}. However, it is when we consider the process of information transfer via entanglement in quantum theory that this relation (arguably equivalence) between entropy and information becomes truly compelling. We begin our exploration of these ideas in Section~\ref{sec:quantum} in the context of quantum measurement\cite{vonNeumann1932mathematical}. In quantum mechanics, entropy is described by the generalisation of the Gibbs-Boltzmann formula in terms of the density matrix of the system\cite{vonNeumann1927theromdynamik}. Here we shall find the tools of decoherence theory\cite{zeh1970interpretation, kubler1973dynamics, zurek1981pointer, zurek1982environment, schlosshauer2007decoherence} invaluable.

In the course of this discussion, we introduce a simple, solvable model for illustrating the nature of information transfer between an observed system and its environment via entanglement. This will serve to more clearly present our development of the concept of entropic time on the basis of information acquisition. The instances of such information gain are associated with the infamous `collapses' of the state vector (or wavefunction) oft discussed in the philosophy of physics. Such collapses may be interpreted as being ‘subjective’, in the sense that this is just how they are perceived by an observer, or ‘objective’ in the sense that alternative possibilities cease to exist.

Later, in Section~\ref{sec:entropic_time}, which shall argue that the notion of entropic time must be the basis for our psychological perception of temporal evolution. We shall also look at some of the deeper implications of subjective and objective collapse theories, which may have real-world consequences for the future of quantum informational technologies. 

%*******************************************************************************************
\section{Entropy and time}\label{sec:thermodynamic}
%*******************************************************************************************
\subsection{Clausius entropy}\label{sec:clausius}
It may seem a bold claim to say that entropy is inexplicable on the basis of classical physics, as there are many intuitive examples of entropy increase that seem perfectly comprehensible. Firstly, let us start with the classical definition of entropy due to Clausius\cite{clausius1865verschiedene}. Denoting entropy by $S$, the change in the entropy of a thermodynamic system is given by

\begin{equation}
dS = \frac{dQ}{T}, \label{eq:Clausius}
\end{equation} 

\noindent where $dQ$ is the heat transfer into or out of the system and $T$ is the absolute temperature.

As an intuitive example of an irreversible process, consider a perfectly insulated box separated into two compartments, each containing an equal number $N$ of identical particles. Let us now suppose that the particles in one side are at a temperature $T_{1} = T+\Delta T$ and those in the other at $T_{2} = T- \Delta T$. If the partition is removed, particles from each side will diffuse into the other, to form a homogeneous mixture with temperature $T$.

This is easily visualised at the microscopic level by imagining the particles in each side to have random positions and velocities, subject to the constraint that the average kinetic energy on either side is equal to $3k_{B}T_{i}/2$, where $k_{B}$ is Boltzmann's constant and $T_{i} = T \pm \Delta T$. Once the partition is removed, those particles that had been heading towards it will now sail through to the other side, after which, all particles are free to bounce throughout the whole container. After a short amount of time, the particles will have mixed well enough that the average energy on either side is now $3k_{B}T/2$ and, at any given time, there will always be roughly $N$ particles on either side.

Macroscopically, the same final state could have been obtained via solution of a text-book constant volume problem whereby each side had been put into thermal contact with a heat bath at temperature $T$. The result for the entropy increase is then found to be

\begin{equation}
\Delta S = \frac{3}{2}Nk_{B}\log\frac{T^{2}}{T^{2} - \Delta T^{2}} > 0. \nonumber
\end{equation}  

\noindent This is then also the increase in entropy via the irreversible process of microscopic diffusion. Since we assume the box to be perfectly isolating, no energy has been transferred to or from the environment of the box during this diffusion. Hence, we see that the total entropy increase of the system and its environment is positive.

This very elementary example appears to indicate that the entropy increases with time. However, time has not appeared explicitly anywhere in our analysis. It \emph{would} appear, however, if we considered this process on the microscopic scale. From the point of view of classical mechanics, we would imagine each particle with an initial position at $t=0$ of $\vec{r}_{i}(0)$ and velocity $\vec{v}_{i}(0)$, all of which would be randomly distributed (subject again to energy conservation). For simplicity, let us suppose that the particles do not interact, so that the magnitudes of their velocities stays constant. At a later time $t$, the new positions will be given by  $\vec{r}_{i}(0) + \vec{v}_{i}t$. Allowing the system to evolve deterministically, we would soon arrive in the homogeneously mixed state with its higher state of entropy. 

Note that mathematically, $t$ is just a parameter that we vary. Associating this with our concept of time is an interpretation of the theory. To make this clear, we shall often refer to $t$ as the \emph{parametric time}.

An important point to note is that the observed increase in entropy just described in microscopic terms is not a consequence of the parametric time $t$ increasing in the positive direction. Macroscopically, the exact same state of affairs would be arrived at if $t$ ran in the negative direction, so that `subsequent' positions of the particles were given by $\vec{r}_{i}(0) - \vec{v}_{i}t$. In other words, the increase in entropy is insensitive to which direction time - that is, the parameter $t$ - is running in. What \emph{is} significant is that we start the system with certain low entropy initial conditions and that, irrespective of the direction we take $t$ to be evolving in, there is a higher entropy state that it naturally evolves into. 

In terms of a unitary theory, however, the final high entropy state would evolve just as naturally into the lower entropy state. The argument for why we do not see this happening in practice is often given in terms of \emph{probability} - the higher entropy states are `far more likely', because there are so many more microscopic states corresponding to the same macroscopic state.

This leads us straight into the work of Boltzmann and Gibbs\cite{boltzmann1866mechanische,jaynes1965gibbs}. However, before venturing down this path, in is salient to note that probability plays no part in the classical mechanics Boltzmann would have been working with. According to classical mechanics, trajectories evolve deterministically - chance has nothing to do with it. Probability comes into the picture purely through our inability to have complete information about a physical system, so that we must resort to statistics to make up for our lack of omniscience. 

\subsection{Gibbs-Boltzmann entropy}\label{sec:GibbsBoltz}
In statistical mechanics, the entropy of a system is defined in terms of the probability of finding it in a particular micro-state (e.g. a particular configuration of the positions and momenta of the constituent particles) corresponding to a given macroscopic state. By `macroscopic state' we mean a state that is characterised in terms of macroscopic variables such as temperature, volume and pressure. If each microscopic state has a probability $p_{i}$, then the Gibbs formula for the entropy\cite{jaynes1965gibbs} is

\begin{equation}
S = -k_{B}\sum_{i}^{\Omega}p_{i}\log p_{i}. \label{eq:Gibbs}
\end{equation}

\noindent where $\Omega$ is the number of micro-states. In the case where all these probabilities are equal, i.e. $p_{i} = 1/\Omega$, this simplifies to the Boltzmann entropy equation\cite{boltzmann1866mechanische}

\begin{equation}
S = k_{B}\log \Omega. \label{eq:Boltzmann}
\end{equation}

We may regain the Clausius expression for entropy from \eqref{eq:Gibbs} by substituting in the Boltzmann factor for the probability of an energy state

\begin{equation}
p_{i} = Z^{-1}e^{-E_{i}/(k_{B}T)}. \label{eq:Boltz_factor}
\end{equation}

\noindent Here, $E_{i}$ is the energy of the $i$th state and $Z$ is the partition function

\begin{equation}
Z = \sum_{i}e^{-E_{i}/(k_{B}T)}. \label{eq:partition}
\end{equation}

Following the argument given in Appendix~\ref{sec:GibbsClausius} (which assumes that the number of particles in the system remains constant), we find that an infinitesimal change in the entropy is given by

\begin{equation}
dS = \frac{\expt{dQ}}{T}, \nonumber
\end{equation}

\noindent where $\expt{dQ}$ is the expectation value of the work done on the system. So, by assuming that the probability of a state is given by the Boltzmann factor, we can establish a connection between the probabilistic description and Clausius' thermodynamic concept of entropy.

%*******************************************************************************************
\section{Entropy and information}\label{sec:information}
%*******************************************************************************************

The work of Boltzmann and Gibbs shows how entropy becomes a concept about statistical ensembles. The greater the number of possible micro-states corresponding to a given macro-state, the higher its entropy. In this picture, an intuitive explanation of why entropy tends to increase is because it does so on the basis of probability: there are far more possible high entropic states than there are low ones, so that the probability that a system evolves into a low entropy state is very low.

Whilst having an intuitive appeal, this probabilistic explanation remains inexplicable in terms of classical mechanics. Under Newtonian laws of motion, there is no `probability' for a system to evolve from one state to another: given a system in an initial state at time $t=t_{0}$, it evolves into one and only one system at $t=t_{1}$. This means that there is only one possible micro-state for the system at any time and that, on the basis of the Boltzmann formula \eqref{eq:Boltzmann}, its entropy must be zero at all times. On the other hand, if the system \emph{were} to change into a more entropic state on the basis of probability, this would evidently not be an evolution arising out of the variation of the parametric time under classical dynamics. 

\subsection{Maxwell's Demon}
A defence of the statistical description of entropy may be made in terms of its utility for dealing with our lack of detailed knowledge of the microscopic state of a system. The use of probability then seems quite justified. However, the penalty for this is that entropy must now be seen as a measure of the information, or rather lack of it, that we have about the system. From the point of view of classical mechanics, this seems very odd. Entropy was introduced by Clausius as a \emph{state variable} of the system: something objective and measurable. Information about the system, on the other hand, would appear  (before the advent of quantum mechanics) to be something subjective. This problem seems to have been known to Maxwell, who invoked his famous `Demon' to exemplify it\cite{maxwell1891theory}.

The details of Maxwell's thought experiment are well-known and we can use our earlier example of the homogeneously mixed gas in a box to discuss it. Suppose we were to place a partition back into the box with a microscopic hatch in it, that may be opened and closed (with negligible dissipation of energy) on command to allow a single gas particle to pass from one side to the other. Maxwell then envisioned a demon, possessed of complete knowledge of all the positions and momenta of all particles in the box. 

With this information, the demon could operate the hatch to only allow fast travelling particles from one side to pass into the other. If it chose, it could also allow slow moving particles to pass in the other direction. In doing so, the demon could manipulate the transfer of energy from one side to the other so raising the temperature in one compartment whilst lowering it in the other - a process in direct violation of the Second Law of Thermodynamics. Moreover, once accomplished, the demon could use the greater pressure of the gas on the hotter side to do work by, say, allowing the partition to move and attaching a load to it.

\subsection{Shannon entropy}
Many attempts at exorcising the demon rely on arguments about the entropy cost of storing or erasing the information about the system and pointing out that the total entropy must apply to both system and demon~\cite{landauer5dissipation,bennett2003notes}. Since these arguments are often made on the basis of classical mechanics, we shall not detour to consider them here, as we are fundamentally interested in the quantum mechanical description of Nature. What Maxwell's Demon does serve to highlight, however, is the profound connection between entropy and information. This connection becomes even more compelling in the light of Shannon's introduction of entropy\cite{shannon1948mathematical,shannon2001mathematical} in the field of information theory. 

Shannon was concerned with the problem of characterising the information content of a stream of data over a communication channel, although the concepts he developed are of far greater applicability. In particular, his work addresses the general question of how we might quantify the information $I$ gained on observing a particular event $i$. 

Suppose that the probability of this observation is $p_{i}$. Now, if $p_{i} = 1$, then the observation gives us no new information, since we knew it was going to happen. In this case, we would require $I(1) = 0$. On the other hand, the smaller the probability, the more information it gives us, since an unlikely event more profoundly revises our expectations. This means that $I(p_{i})$ should decrease monotonically as $p_{i}$ increases. We would also like the information provided by mutually exclusive events to be additive, so that for events $i$ and $j$, $I(p_{i}p_{j}) = I(p_{i}) + I(p_{j})$. These requirements are all met if we define our basic unit of information to be

\begin{equation}
I(p_{i}) = -\log_{2}p_{i}. \label{eq:basic_info}
\end{equation}

\noindent Here, the logarithm is in base 2 since Shannon was concerned with binary digits (bits) of information. Later, we shall adopt natural logarithms, giving information in units of `nats'. Shannon's entropy measure is then defined to be the expectation value of the information provided by a set of events with probabilities $p_{i}$. That is

\begin{equation}
S = -\sum_{i}p_{i}\log_{2} p_{i}. \label{eq:Shannon}
\end{equation}

Shannon's expression for entropy is almost identical to the Gibbs formula \eqref{eq:Gibbs} save for the factor of $k_{B}$ and the use of base 2 for the logarithm. In fact, we may remark that the appearance of $k_{B}$ in the thermodynamic formulas is a legacy of measuring temperature in different units to energy. Since temperature is a measure of average energy, there is no \emph{a priori} reason why we should treat it as being a different physical dimension. If we \emph{were} to measure temperature and energy in the same units, thermodynamic entropy would become dimensionless, as it is in information theory.

Given that we can make the thermodynamic entropy and informational entropy dimensionally equivalent, there is no conceptual problem with discussing them on the same footing. Indeed, the discussions above have suggested that thermodynamic entropy is dependent on the information we have about the system. 

%*******************************************************************************************
\section{Entropy and quantum measurement}\label{sec:quantum}
%*******************************************************************************************
\subsection{von Neumann entropy}
The quantum mechanical description of entropy is given by von Neumann's formula\cite{vonNeumann1927theromdynamik} in terms of the density matrix of a system $\rho$

\begin{equation}
S = -\mathrm{Tr}\left[\rho\log\rho\right] = -\sum_{i} \lambda_{i}\log\lambda_{i}, \label{eq:von_Neumann}
\end{equation}

\noindent where `$\mathrm{Tr}$' stands for the trace operator and the $\lambda_{i}$ are the eigenvalues of $\rho$. The analogy between this and either the Gibbs-Boltzmann or Shannon entropies is easy to see if these eigenvalues turn out to be probabilities (which they invariably do). The advantage of using this formalism in dealing with open quantum systems is that we can consider both \emph{pure-state} density matrices formed from the state vector of the system $\ket{\Psi}$ and \emph{mixed-state} matrices that represent a classical ensemble of states. For a pure-state, we have simply

\begin{equation}
\rho_{pure} = \ket{\Psi}\bra{\Psi}, \label{eq:pure_state}
\end{equation}

\noindent so that, if this were represented in a basis of eigenvectors $\ket{\phi_{i}}$ via

\begin{equation}
\ket{\Psi} = \sum_{i} a_{i}\ket{\phi_{i}}, \nonumber
\end{equation}

\noindent we would have

\begin{equation}
\rho_{pure} = \sum_{ij}a_{i}a_{j}^{*}\ket{\phi_{i}}\bra{\phi_{j}}. \label{eq:pure}
\end{equation}

\noindent This has off-diagonal elements known as \emph{coherences} and are related to the interference phenomena characteristic of quantum mechanics.

On the other hand, we might wish to model an ensemble of states in which we assign a classical probability $p_{i}$ to each state. In this case, we would have

\begin{equation}
\rho_{mix} = \sum_{i}p_{i}\ket{\phi_{i}}\bra{\phi_{i}}. \label{eq:mixed}
\end{equation}

\noindent Here, there are no coherences although, according to the Born Rule~\cite{born1926quantenmechanik} for interpreting the squared moduli of quantum amplitudes, the diagonal elements are the same as in the pure state case, i.e. $p_{i} = \sqmod{a_{i}}$. For the mixed-state density matrix, the eigenvalues are just these probabilities and so the von Neumann expression \eqref{eq:von_Neumann} just reproduces the Gibbs entropy (except for the Boltzmann constant of proportionality).

In the case of a pure-state density matrix, however, it can be shown that the von Neumann entropy is identically zero at all times\footnote{See Appendix~\ref{sec:zero_entropy}.}. We may interpret this is a concrete demonstration of the consequence of deterministic evolution, discussed earlier in the case of classical mechanics. There we argued that the deterministic evolution of a classical system could never lead to an increase in the Gibbs-Boltzmann entropy, since a given configuration at time $t_{0}$ will evolve into one and only one configuration at time $t_{1}$. Despite the probabilistic interpretation of quantum mechanics via the Born rule, the evolution of quantum amplitudes is just as deterministic as classical dynamics and so the same phenomenon of the `conservation of zero entropy' pertains. 

It may appear, then, that our hopes of understanding entropy from a quantum mechanical perspective have been scuppered. In fact, the root of the problem is a far more wide-ranging. According to the Born Rule, we may interpret quantum mechanics statistically by interpreting the squared moduli of amplitudes as probabilities (or probability density functions) for observing the physical quantities that those states represent. However, there is nothing in the formal apparatus of conventional quantum mechanics that allows us to go from a probability to an actual, unique observation. This process may be described as a `projective measurement' or, according to some interpretations, `wave function collapse'. Such processes are inherently irreversible and, whilst not predicted or explained by the theory, evidently happen. Such irreversibility is the hall mark of increasing entropy.

The difficulties of dealing with projective measurement are an intrinsic part of the well-known \emph{measurement problem}. A great deal of insight into this problem has been gleaned through the formal development of decoherence theory\cite{zeh1970interpretation, kubler1973dynamics, zurek1981pointer, zurek1982environment, schlosshauer2007decoherence} and its interpretation. One of the most appealing aspects of these approaches is that they do no harm to the existing structure of quantum theory, so we do not have to adopt any radically new axioms or postulates. It also emphasises information as a fundamentally physical concept.

\subsection{The relative state formalism}
A scheme of ideal quantum measurement was introduced by von Neumann\cite{vonNeumann1932mathematical}, in which a system under study interacts with a measuring apparatus without changing its state. The apparatus (or, more generally, the environment of the system) then encodes information about the system `pointer' state\cite{zurek1981pointer}. Symbolically, this can be represented as

\begin{equation}
\ket{\phi_{i}}\ket{A_{0}} \to \ket{\phi_{i}}\ket{A_{i}}, \label{eq:von_Neumann1}
\end{equation}

\noindent where $\ket{\phi_{i}}$ is an eigenstate of the system and $\ket{A_{0}}$ is the `ready state' of the apparatus. More precisely, this evolution may be specified by the operation of a unitary operator $U(t) = e^{-iHt/\hbar}$, where $H$ is the total Hamiltonian of the system, apparatus and the interaction between them. 

In general, the system will be in a superposition of eigenstates. Then, due to the linearity of the Hamiltonian, we would have

\begin{equation}
\sum_{i}a_{i}\ket{\phi_{i}}\ket{A_{0}} \to \sum_{i}a_{i}\ket{\phi_{i}}\ket{A_{i}}. \label{eq:von_Neumann2}
\end{equation}

\noindent An important point to note about the expression on the right is that this represents an \emph{entangled} state. That is, it cannot be factored into a product of system and apparatus states. It is via this entanglement that information about the state of the system is transferred to the apparatus. For instance, if we observe the apparatus (a macroscopic object) in the state $\ket{A_{i}}$ we now know that the system is in state $\ket{\phi_{i}}$. At the same time, however, all information about the other states, represented by the amplitudes of these states, is now lost. This net loss of information may be interpreted as an increase in entropy.

\subsubsection{A bipartite open system}\label{sec:bipartite_open}
Before looking at how we may calculate this increase in entropy, let us first review the dynamics of open systems. For simplicity, we shall just consider a bipartite system consisting of a subsystem $\S$ and its environment $\Env$, defined on Hilbert spaces $\mathcal{H}_{\S}$ and $\mathcal{H}_{\Env}$ respectively. The Hilbert space of the total system will then be the tensor product $\mathcal{H} = \mathcal{H}_{\S}\otimes\mathcal{H}_{\Env}$.

The Hamiltonian of the system will be of the form

\begin{equation}
H = H_{\S} + H_{\Env} + H_{\S\Env}, \label{eq:H_form}
\end{equation}

\noindent where $H_{\S\Env}$ defines the interaction between $\S$ and $\Env$. If $\{\ket{\f_{i}}\}$ and $\{\ket{\E_{j}}\}$ are complete, orthonormal sets of eigenvectors spanning $\mathcal{H}_{\S}$ and $\mathcal{H}_{\Env}$ respectively, then a general state of the system may be written as

\begin{equation}
\ket{\Psi(t)} = \sum_{ij} a_{ij}(t)\ket{\f_{i}\E_{j}}, \label{eq:Psi_general}
\end{equation}

\noindent where we have used the short-hand $\ket{\f_{i}\E_{j}} = \ket{\f_{i}}\otimes\ket{\E_{j}}$ for the tensor product of system and environment states.

We can re-write \eqref{eq:Psi_general} using Everett's relative state formulation\cite{everett1956theory} as

\begin{equation}
\ket{\Psi(t)} = \sum_{i}a_{i}\ket{\f_{i}\R_{i}(t)}, \label{eq:Psi_rel}
\end{equation}

\noindent where we have defined the \emph{relative state} of $\ket{\f_{i}}$ in $\Env$ as

\begin{equation}
\ket{\R_{i}(t)} = \sum_{j}\frac{a_{ij}(t)}{a_{i}}\ket{\E_{j}}. \label{eq:R}
\end{equation}

\noindent Note that, in general, the $\ket{\R_{i}(t)}$ will \emph{not} be orthogonal to one another, although in cases where they become orthogonal, such relative states could represent the apparatus states of \eqref{eq:von_Neumann2}. As we shall see, the degree of orthogonality of these relative states is indicative of the degree of entanglement between the system and environment.

Whether or not the relative states do become orthogonal will be determined by the time evolution of the inner products

\begin{equation}
\iprod{\R_{i'}(t)}{\R_{i}(t)} = \sum_{j}\frac{a_{i'j}^{*}(t)a_{ij}(t)}{a_{i'}^{*}a_{i}}. \label{eq:iprod_general}
\end{equation}

\noindent As an immediate observation, we note that, since $\sqmod{a_{i}} = \sum_{j} \sqmod{a_{ij}}$ is the marginal probability of for the system being found in the state $\ket{\phi_{i}}$, we have $\iprod{\R_{i}(t)}{\R_{i}(t)} = 1$, as expected. We may further note that, if we start the total system in a tensor product of system and environmental states by putting $a_{ij}(0) = a_{i}b_{j}$, then at $t = 0$, we have

\begin{equation}
\iprod{\R_{i'}(0)}{\R_{i}(0)} = \sum_{j}\sqmod{b_{j}} = 1. \nonumber
\end{equation}

\noindent In other words, if we start things off in a tensor product, then at $t=0$ all relative states are equivalent. At this point, there is no entanglement between the system and environment. Any orthogonality between the relative states can only emerge with the temporal evolution of the total system. We shall consider possible conditions for orthogonality later in Section~\ref{sec:DEM} in terms of a simple, soluble model.

\subsubsection{The reduced density matrix}\label{sec:reduced_density}
Let us see how these ideas connect with information and entropy. A standard mathematical tool of decoherence theory is the \emph{reduced density matrix}. This is obtained from the total density matrix by tracing out the degrees of freedom of the environment to obtain the reduced density matrix of the system alone (or vice versa if one wishes). That is, denoting the system reduced density matrix by $\rho_{\S}$, 

\begin{equation}
\rho_{\S} = \sum_{j}\iprodM{\E_{j}}{\rho}{\E_{j}}, \label{eq:red_rho}
\end{equation}

\noindent where $\rho = \oprod{\Psi}{\Psi}$ is the pure-state density matrix of the total system.

Using \eqref{eq:Psi_general}, for the general state, we obtain the reduced density matrix

\begin{equation}
\rho_{\S}(t) = \sum_{ii'j}a_{ij}(t)a_{i'j}^{*}(t)\ket{\f_{i}}\bra{\f_{i'}}. \label{eq:red_rho_rel0}
\end{equation}

\noindent However, on inspection of \eqref{eq:iprod_general} for the inner product of relative states, we see that this may be written

\begin{equation}
\rho_{\S}(t) = \sum_{ii'}a_{i}a_{i'}^{*}\iprod{\R_{i'}(t)}{\R_{i}(t)}\ket{\f_{i}}\bra{\f_{i'}}. \label{eq:red_rho_rel}
\end{equation}

\noindent Evidently, the inner products of different relative states sit on the off-diagonal elements of the reduced density matrix (on the diagonal elements $\iprod{\R_{i}}{\R_{i}} \equiv 1$). If the relative states become orthogonal as $\ket{\Psi}$ evolves, these off-diagonal elements will disappear, leaving just the probabilities $p_{i} = \sqmod{a_{i}}$ along the diagonal. In other words, the reduced density matrix may become mathematically equivalent to a mixed density matrix, for which we would have a finite entropy via the von Neumann formula \eqref{eq:von_Neumann}.

Consider the particular case in which we start the total system off at $t=0$ in a tensor product, so we have $\iprod{\R_{i'}(0)}{\R_{i}{0}} = 1$ for all $i', i$. Initially, then, we have

\begin{equation}
\rho_{\S}(0) = \sum_{ii'}a_{i}a_{i'}^{*}\ket{\f_{i}}\bra{\f_{i'}}. \label{eq:red_pure}
\end{equation}

\noindent This is just the density matrix of the isolated system. Being a pure density matrix, it will have an entropy of zero. 

Now let us suppose that as the system evolves, all the off-diagonal elements disappear. The eigenvalues of $\rho_{\S}(t)$ would then be equal to the diagonal elements and we would have

\begin{equation}
\lim_{\iprod{R_{i'}}{R_{i}}\to \delta_{i'i}} S(\rho_{\S}) = -\sum_{i} \sqmod{a_{i}}\log\sqmod{a_{i}}. \label{eq:von_Neumann_inf}
\end{equation}

\noindent This is equal to the entropy of a classical ensemble of micro-states, each with a probability of $\sqmod{a_{i}}$. 

From the point of view of a particular relative state $\ket{\R_{i}}$, the system is now in a definite state $\ket{\f_{i}}$, which we can regard as a transfer of information to $\ket{\R_{i}}$. Specifically, in accordance with \eqref{eq:basic_info}, we shall assert that the information transferred to the relative state $\ket{\R_{i}}$ in this limit is

\begin{equation}
I(\R_{i}) = -\log\sqmod{a_{i}} \label{eq:I_Ri}
\end{equation}

\noindent (in units of `nats' since we are using natural logarithms). At the same time, however, information about the other states, including the original amplitude of $\ket{\f_{i}}$ has now been lost, so \eqref{eq:von_Neumann_inf} is interpreted as a measure of the \emph{loss} of information. We may then define the net change in information with respect to the $k$th relative state, in the limit of the relative states becoming orthogonal, as

\begin{equation}
\lim_{\iprod{R_{i'}}{R_{i}}\to \delta_{i'i}}\Delta S_{k}(\rho_{S}) = -\left(\sum_{i} \sqmod{a_{i}}\log\sqmod{a_{i}} - \log\sqmod{a_{k}}\right). \label{eq:Delta_Sk}
\end{equation}

We may note that if we were to choose the system basis states $\ket{\f_{i}}$ to be eigenstates of $\rho_{S}$ with eigenvalues $\lambda_{i}$, we would have, from \eqref{eq:red_rho_rel}, 

\begin{align}
\rho_{\S}\ket{\f_{k}} &= \sum_{i}a_{i}a_{k}^{*}\iprod{\R_{k}}{\R_{i}}\ket{\f_{i}}, \nonumber \\
&= \lambda_{k}\ket{\f_{k}}. \nonumber
\end{align}

\noindent This \emph{requires} that $\iprod{\R_{k}}{\R_{i}} = \delta_{ki}$ and so $\lambda_{k} = \sqmod{a_{k}}$, enforcing the required form of \eqref{eq:von_Neumann2} for ideal quantum pre-measurement. This is known as the \emph{Schmidt decomposition}~\cite{schmidt1908theorie}. Whilst the practicality of this decomposition is limited due to the enforced time dependent nature of the eigenvectors (since $\rho_{\S}$ is time dependent), this result may be used to derive\footnote{See Appendix~\ref{app:entropy_entanglement}.} a remarkable theorem for bipartite systems that 

\begin{align}
S(\rho_{S}) = S(\rho_{E}). \label{eq:bipartite_entropy_theorem}
\end{align}

Here $\rho_{E}$ is the reduced density matrix of the environment formed by tracing out the degrees of freedom of the system. This result is by no means obvious from the form of \eqref{eq:red_rho_rel0} for $S(\rho_{S})$ and its counterpart for $S(\rho_{E})$. This equivalence allows us to think of the entropy exchange as a mutual feature of the subsystems of a bipartite system and is known in the literature as the \emph{entropy of entanglement}.

\subsubsection{Evolution of $\rho_{S}$}
It is straightforward to show that without interaction, a bipartite system can never become entangled and thus, no information can be transferred between its components. Setting $H_{\S\Env} = 0$ in \eqref{eq:H_form} and starting the system off in a tensor product of system and environmental states, the temporal evolution is given by

\begin{align}
\ket{\Psi(t)} &= \rme^{-\rmi\left(H_{\S} + H_{\Env}\right)t/\hbar}\sum_{i} a_{i}\ket{\f_{i}}\otimes \sum_{j} b_{j}\ket{\E_{j}}, \nonumber \\
&= \sum_{i} a_{i}\rme^{-\rmi H_{\S}t/\hbar}\ket{\f_{i}}\otimes \sum_{j} b_{j}\rme^{-\rmi H_{\Env}t/\hbar}\ket{\E_{j}}, \nonumber
\end{align}

\noindent which clearly remains in a tensor product at all times.

The dynamic orthogonalisation of relative states, picking out pointer states of the system, therefore requires interaction between subsystem and environment. As we shall also see in Section~\ref{sec:DEM} in the context of a soluble solution, the effectiveness of this orthogonalisation depends to some degree on the distribution of environmental states. It is also clear from the discussion of the Schmidt decomposition in the previous section that certain sets of subsystem eigenstates have relatives states that more readily become orthogonal than others. Put the other way around, the features of the environmental interaction and distribution of states that most readily lead to orthogonal relative states naturally select out preferred subsystem eigenstates. Zurek has dubbed this \emph{environment-induced superselection}~\cite{zurek1981pointer,zurek1982environment}. 

As an illustration of this effect, choosing subsystem states $\ket{\f_{i}}$ and environmental states $\ket{\E_{j}}$ that diagonalise the interaction matrix $H_{\S\Env}$, the time derivative of the elements of the reduced density matrix $\rho_{\S}$ is found to be\footnote{See Appendix~\ref{app:drho_dt}\label{test}}

\begin{equation}
\frac{d\rho_{ij}}{dt} = -\frac{\rmi}{\hbar}\iprodM{\f_{i}}{\left[H_{\S},\rho_{\S}\right]}{\f_{j}} +  \rho_{ij}\frac{d}{dt}\log\iprod{\R_{j}}{\R_{i}}, \label{eq:drho_dt}
\end{equation} 

\noindent where $\rho_{ij} = \iprodM{\f_{i}}{\rho_{\S}}{\f_{j}}$ and $\left[x,y\right] = xy - yx$ is the commutation operator. Without the last term, this would be the Liouville–von Neumann equation for the unitary evolution of the density matrix. The term involving the relative state inner products $\iprod{\R_{j}}{\R_{i}}$ introduces a non-unitary term that, in general, acts to suppress the off-diagonal elements. The dynamics of the $\iprod{\R_{j}}{\R_{i}}$ are in turn determined by the Hamiltonians $H_{\Env}$, $H_{\S\Env}$\footnote{See Section~\ref{sec:DEM}} and the distribution of the environmental states\footnote{\emph{ibid}}.

\subsection{A diagonal energy model}\label{sec:DEM}
\subsubsection{A bipartite system}
As an illustrative example of the relative state formulation, we describe a model system for which the time evolution can be solved in closed form. To begin with, we shall just consider the bipartite system described by \eqref{eq:H_form} and take the interaction matrix $H_{\S\Env}$ to be diagonal in the representation $\ket{\f_{i}\E_{j}}$, where $\ket{\f_{i}}$ and $\ket{\E_{j}}$ are energy eigenvectors of $H_{\S}$ and $H_{\Env}$ respectively. $H_{\S\Env}$ then has the form

\begin{equation}
H_{\S\Env} = \sum_{ij}\hbar\omega_{ij}\ket{\f_{i}}\bra{\f_{i}}\otimes\ket{\E_{j}}\bra{\E_{j}}. \label{eq:H12}
\end{equation}

\noindent Here, we identify the energies associated with each component of the total system via the consistent use of indices, always using $i$ to index subsystem states and $j$ for the environmental states, as in the equation above.

Denoting the eigenvalues of $H_{\S}$ and $H_{\Env}$ by $\hbar\omega_{i}$ and $\hbar\omega_{j}$ respectively, the eigenvalues of the total system are $E_{ij} = \hbar(\omega_{i} + \omega_{j} + \omega_{ij})$. However, the notation is simplified by moving to the interaction picture via the transformation $\ket{\Psi(t)} \to e^{\rmi\left(H_{\S}+H_{\Env}\right)t/\hbar}\ket{\Psi(t)}$ (the interaction matrix, being diagonal in the energy eigenvectors, is unaffected by this  transformation). We then need only concern ourselves with the interaction energies.

If we now start the system in a tensor product of system and environment states, so that $a_{ij}(0) = a_{i}b_{j}$, the time evolution of the system is then given by

\begin{equation}
\ket{\Psi(t)} = \sum_{ij} a_{i}b_{j}\rme^{-\rmi\omega_{ij}t}\ket{\f_{i}\E_{j}}. \label{eq:model_general}
\end{equation}

\noindent In this model we see that, even though the coefficients of the subsystem states do not change, the subsystem becomes entangled with the environment via the temporal phase terms.

In terms of the relative state expansion, we now have

\begin{equation}
\ket{\Psi(t)} = \sum_{i}a_{i}\ket{\f_{i}\R_{i}(t)}, \label{eq:Psi_rel2}
\end{equation}

\noindent and

\begin{equation}
\ket{\R_{i}(t)} = \sum_{j}b_{j}\rme^{-\rmi\omega_{ij}t}\ket{\E_{j}}. \label{eq:R2}
\end{equation}

\noindent Again, at $t = 0$, all relative states are equal. We may denote this state by 

\begin{equation}
\ket{\R_{0}} = \sum_{j}b_{j}\ket{\E_{j}}. \label{eq:R0}
\end{equation}

\noindent In terms of the von Neumann scheme of measurement, $\ket{\R_{0}}$ then represents the `ready state' of the apparatus or environment.

\subsubsection{Orthogonality of the relative states}\label{sec:DEM_ortho}
Taking the inner product of two relative states given by \eqref{eq:R2}, we have

\begin{equation}
\iprod{\R_{i'}(t)}{\R_{i}(t)} = \sum_{j}\sqmod{b_{j}}\rme^{\rmi(\omega_{i'j} - \omega_{ij})t}. \label{eq:R_inner}
\end{equation}

\noindent Due to the normalisation of the environmental states, this yields unity for all states at $t = 0$ and at all times for $i' = i$. For $i' \ne i$, the behaviour of the inner product will depend on both the initial superposition of the environmental states and the variation of the interaction energy differences with $j$. If, for example, all the interaction energies were equal, then the relative states could never become orthogonal to one another. On the other hand, if (for example) the frequency differences $\omega_{i'j} - \omega_{ij}$ vary linearly with $j$, then \eqref{eq:R_inner} has the form of a Fourier expansion.

Suppose further that all the coefficients $b_{i}$ were of equal magnitude up to a certain energy, representing a step function in energy. The Fourier transform of this in the continuum limit is a sinc function, which approaches zero as $t\to\pm\infty$. In this case, the relative states \emph{do} become orthogonal as the system evolves on a time scale $\tau = 2\pi/\Omega$, where $\hbar\Omega$ is the range of the interaction energies spanned by the environmental states with $b_{j} \ne 0$.\footnote{See Appendix~\ref{app:sinc}}. 

Alternatively, we may assume a thermal distribution for the environmental states of the form $\sqmod{b_{\w}} = \tau\rme^{-\w\tau}$, where $\tau = \hbar/(k_{B}T)$ and $k_{B}T$ is the thermal energy. Assuming that $\w_{i'j} - \w_{ij} \propto \w$ and absorbing any constant of proportionality into the time units, \eqref{eq:R_inner} then takes the form\footnote{See Appendix~\ref{app:thermal}}

\begin{equation}
\iprod{\R_{i'}(t)}{\R_{i}(t)} = \frac{\rme^{\rmi\theta(t)}}{\left(1 + t^{2}/\tau^{2}\right)^{1/2}}, \nonumber 
\end{equation}

\noindent where $\theta(t) = \tan^{-1}(t/\tau)$. Again, this approaches zero asymptotically for $t\to\pm\infty$.

We should bear in mind, however, that this asymptotic behaviour only strictly pertains for a continuous integral over $\omega_{i'j} - \omega_{ij}$. For a sum over discrete terms, we would generally have some sort of oscillatory behaviour, albeit, perhaps, with extremely long periods. Typically, though, the continuum approximation will be a good one, since the Hilbert space of the environment will be vast and the energy levels closely spaced or quite possibly continuous.

%*******************************************************************************************
\section{Entropic time}
%*******************************************************************************************
\subsection{Temporal boundary conditions}
In discussion of the evolving entanglement of a subsystem with its environment, we have taken the state at $t=0$ to be a tensor product of $\S$ and $\Env$ states for which there is no entanglement. Such an initial state represents a significant temporal boundary condition. In particular, the entropy of the reduced system matrix \emph{will be zero}. In the presence of some kind of interaction between the two, the temporal evolution of the system may then lead to $\S$ and $\Env$ becoming entangled, leading to a non-zero entropy of entanglement. 

It is pertinent to appreciate that, due to the unitary nature of the operator $U = \rme^{-\rmi Ht/\hbar}$ determining this interaction, this entanglement will emerge whether we take $t$ to proceed in the positive or negative direction. Moreover, since the only difference between the solutions will be a phase factor of order unity, the entropy of entanglement, being related to the squared moduli of amplitudes, will be the same in either case. An explicit example of this may be seen from the soluble solution of Section~\ref{sec:DEM}.

This may be compared to the classical example of dispersing particles discussed earlier in Section~\ref{sec:clausius}. There, it was argued that the increase in thermodynamic entropy was not so much due to a directionality in time but that we were starting in a particularly ordered state, which then naturally evolved into a higher entropy state. That is, the increase in entropy is relative to the initial boundary condition. In the quantum mechanical case, this is the non-entangled state of affairs.

One might then ask how reasonable it is to assume that we can ever start under such specialised conditions, given how readily systems interact with each other. In fact, assuming that relative states do readily become orthogonal, we might argue that such specialised conditions are occurring all the time. Every component of the superposition of \eqref{eq:von_Neumann2} for the scheme of quantum pre-measurement is a tensor product of subsystem and environmental states. The actuality of any observation is taken to be the `collapse' of all other possibilities to one of these components.

\subsection{Information gain on `collapse'}\label{sec:info_gain}
Instead of assuming that a bipartite system starts off in a tensor product, let us assume that at some time $t_{0}$, it has evolved to a superposition of states

\begin{equation}
\ket{\Psi(t_{0})} = \sum_{i}a_{i}(t_{0})\ket{\f_{i}\R_{i}(t_{0})}, \label{eq:Psi_t1}
\end{equation}

\noindent where the $\ket{\f_{i}}$ are taken to be time-independent vectors that, along with environmental states $\ket{\E}$, diagonalise the interaction matrix. Moreover, we assume that the relative states have become orthogonal so that $\iprod{\R_{i'}(t_{0})}{\R_{i}(t_{0})} = \delta_{i'i}$, meaning that the system has effectively `collapsed'. At this point, an observer in relative state $\ket{\R_{i}}$ will definitely see the subsystem $\S$ in state $\ket{\f_{i}}$, which we may interpret via \eqref{eq:I_Ri} as a gain in information of $I_{i}(t_{0}) = -\log \sqmod{a_{i}(t_{0})}$ (in units of nats).

The subsequent time evolution of the joint system might then proceed in any number of ways. One possibility is that our joint system comes into contact with a \emph{larger} environmental system (or perhaps the subsystem only interacted with a subspace of its environment in the first place). Conceivably, this could just erase the information stored in the orthogonal relative states of \eqref{eq:Psi_t1}. However, if this \emph{did} happen, we, as observers, could not know anything about this, since the information that these states had formed would no longer exist.

Alternatively, we might suppose that the information contained in the relative states became \emph{robustly} stored in the environment, so providing a permanent record. The joint system may then evolve to form \emph{new} relative states with the larger environment, each of which would have evolved from the $\ket{\f_{i}\R_{i}(t_{0})}$ pairs. For example, suppose the joint system now interacts to pick out new system pointer states $\ket{\xi_{i'}}$. Expanding the $\ket{\f_{i}}$ in this new basis, we can put

\begin{equation}
\ket{\f_{i}\R_{i}(t_{0})} = \sum_{i'}a_{i'}^{i}\ket{\xi_{i}\R_{i}(t_{0})}, \nonumber
\end{equation}

\noindent where $a_{i'}^{i} = \iprod{\xi_{i'}}{\f_{i}}$. Representing the larger environment ready state by $\ket{\R_{0}^{i}}$, we might then get the evolution

\begin{equation}
\ket{\f_{i}\R_{i}(t_{0})\R_{0}^{i}} \to \sum_{i'}a_{i'}^{i}\ket{\xi_{i}\R_{i}(t_{0})\R_{i'}^{i}(t_{1})}. \nonumber
\end{equation}

Here, we suppose that at time $t_{1}$, the relative states are once more orthogonal. At this point, the original transfer of information based on the probability $\sqmod{a_{i}}$ is supplemented by the additional probability $\sqmod{a_{i'}^{i}}$. Hence the total information transferred to the environment is now 

\begin{align}
I_{i'}(t_{1}) &= I_{i}(t_{0}) - \log \sqmod{a_{i'}^{i}(t_{1})}, \nonumber \\
& = - \log \left(\sqmod{a_{i}(t_{0})} \sqmod{a_{i'}^{i}(t_{1})}\right). \nonumber
\end{align}

\noindent Further increments of information gain may then occur at other times $t_{n}$ in a similar way. 

Note that we need not make any assumption about the ordering of these $t_{n}$ in terms of the temporal parameter $t$ but rather that there is inherent ordering in terms of information gain 

\begin{equation}
I(t_{n}) > \ldots > I(t_{1}) > I(t_{0}). \label{eq:I_t_order}
\end{equation} 

\noindent This implies the kind of \emph{causal} ordering often associated with our everyday perception of time. Moreover, since this ordering is based on information transfer, it is naturally associated with entropy increase. This then restores the perception of the arrow of time at the level of the relative states that is absent from the unitary evolution at the universal level. 

This is then the basis of our concept of \emph{entropic time}. We now have definitively ordered time instances $t_{0}, t_{1}, \ldots t_{n}$ based on \emph{information gain}. Whilst the definition of such instances is best made in terms of these increments of information, a measure with the dimensions of time could be made in terms of the characteristic time $\tau$ discussed in Section~\ref{sec:DEM_ortho} for the time it takes for the relative states to become orthogonal. Of course, this would then be system dependent. 

\subsection{Effective non-unitarity}
Whilst at the universal level (that is at the level of the \emph{total system}), the dynamics of the total state vector will always be unitary, at the level of the system / relative state tensor products, we see the emergence of an `effective' non-unitarity. 

Under the action of a time operator $U(t) = \rme^{-\rmi Ht/\hbar}$, the product of a system state with an environmental `ready' state will evolve into a superposition of pointer state/relative state pairs. Moreover, this will happen whether we take the parametric time $t$ to be in the positive or negative direction. If the system then becomes entangled, the entropy of entanglement will be the same in either case.

Applying the conjugate transpose of the time operator $U^{\dagger} = \rme^{\rmi Ht/\hbar}$ to this \emph{total} superposition will then take the system back to its original tensor product state. However, applying $U^{\dagger}$ to any \emph{particular} pointer state/relative state pair will \emph{not} do so. At the level of this pair, the tensor product must be viewed as its own initial state and cannot be returned to the initial state of the total system.

Any observation of any physical system always `collapses' the system to one of these pointer state/relative state tensor products. From the point of view of an \emph{objective} collapse interpretation, such as the Ghirardi-Rimini-Weber (GRW) \cite{ghirardi1986unified} scheme of spontaneous localisation, this is a decidedly irreversible and non-unitary process. Under this interpretation, the observed evolution of a system would necessarily consist of a sequence of entropic time steps, as described in the previous subsection.

On the other hand, if we apply a \emph{subjective} collapse interpretation, we suppose that the system only \emph{appears} to behave in this way and that the full superposition of possibilities continues to exist in some sense. We might then suppose that Everett's many worlds interpretation\cite{everett1956theory} might apply, although, as we discuss in Section \ref{sec:entropy-energy-conservation}, this is by no means a necessary conclusion.

Either way, if information about an observation is stored robustly in the environment, further interaction with the environment may encode a \emph{history} of a particular evolutionary path. From `within' such a history, the apparent evolution is no longer unitary. Whilst the history might always be erased at the universal level, no observer could know about this (since the information we be lost). We can only have information about those histories that \emph{have} been robustly stored in the environment.  

\subsection{Entropy and energy}
\subsubsection{Energy expectation values}
Although decoherence generally operates on a much faster time scale than dissipation~\cite{schlosshauer2007decoherence_ch2p93}, in classical thermodynamics, entropy has a deep connection to the availability of energy. In this section, we shall investigate this connection in more detail and in particular reference to the issue of energy conservation. 

Choosing basis vectors $\ket{\f_{i}}$ and $\ket{\E_{j}}$ that diagonalise the interaction matrix, we may write this component of  the Hamiltonian for a bipartite system as we did earlier in \eqref{eq:H12}. These will be the basis vectors that we shall use in the representation of a general state of the system.

The system and environmental Hamiltonians may be written quite generally as

\begin{equation}
H_{\S} = \sum_{i}\hbar\omega_{i}^{\S}\ket{\varphi_{i}}\bra{\varphi_{i}} \label{eq:HS}
\end{equation}

\noindent and

\begin{equation}
H_{\Env} = \sum_{j}\hbar\omega_{j}^{\Env}\ket{\eta_{j}}\bra{\eta_{j}}, \label{eq:HE}
\end{equation}

\noindent where the $\ket{\varphi_{i}}$ and $\ket{\eta_{j}}$ are the energy eigenvectors of $\S$ and $\Env$ with eigenvalues $\hbar\omega_{i}^{\S}$ and $\hbar\omega_{j}^{\Env}$ respectively.

Any coupling between the tensor products $\ket{\f_{i}\R_{i}}$ of the relative state expansion, given by \eqref{eq:Psi_rel}, may only come about through the matrix elements $\iprodM{\f_{i'}\R_{i'}}{H}{\f_{i}\R_{i}}$. Inspecting each of the components of this, we find

\begin{equation}
\iprodM{\f_{i'}\R_{i'}}{H_{\S}}{\f_{i}\R_{i}} = \iprod{\R_{i'}}{\R_{i}}\sum_{i}\iprod{\f_{i'}}{\varphi_{i}}\hbar\omega_{i}^{\S}\iprod{\varphi_{i}}{\f_{i}}, \label{eq:coupling_S}
\end{equation}

\begin{equation}
\iprodM{\f_{i'}\R_{i'}}{H_{\Env}}{\f_{i}\R_{i}} = \delta_{i'i}\sum_{j}\iprod{\R_{i}}{\eta_{j}}\hbar\omega_{j}^{\Env}\iprod{\eta_{j}}{\R_{i}} \label{eq:coupling_E}
\end{equation}

\noindent and

\begin{align}
\iprodM{\f_{i'}\R_{i'}}{H_{\S\Env}}{\f_{i}\R_{i}} &= \delta_{i'i}\sum_{j}\iprod{\R_{i}}{\E_{j}}\hbar\omega_{ij}\iprod{\E_{j}}{\R_{i}}, \nonumber \\
&= \delta_{i'i}\sum_{j}\frac{\sqmod{a_{ij}}}{\sqmod{a_{i}}}\hbar\omega_{ij}. \label{eq:coupling_SE}
\end{align}

The only one of these components that does not necessarily go as $\delta_{i'i}$ is then \eqref{eq:coupling_S} involving the system Hamiltonian. This also becomes proportional to $\delta_{i'i}$ when the relative states become orthogonal. In this case, we can write down an expression for the energy expectation of the total system via $\expt{E} = \iprodM{\Psi}{H}{\Psi}$ as

\begin{align}
\lim_{\iprod{R_{i'}}{R_{i}}\to \delta_{i'i}}\expt{E} &= \sum_{i}\sqmod{a_{i}}\iprodM{\f_{i}\R_{i}}{H}{\f_{i}\R_{i}}, \nonumber \\
&\equiv \sum_{i}\sqmod{a_{i}}\expt{E_{i}}, \label{eq:cons_exptE}
\end{align}

\noindent where we have defined $\expt{E_{i}}$ to be the energy expectation value for $i$th relative state product $\ket{\f_{i}\R_{i}}$.

\subsubsection{Entropy and energy conservation}\label{sec:entropy-energy-conservation}
Up to this point, we have taken the view that no choice between subjective and objective collapse theories needs to be made for practical considerations. However, where philosophical contention between these views has arisen, it is often fought out over the issue of energy conservation. In particular, it is often argued that subjective collapse interpretations violate this principle. This is perhaps ironic, since as we shall argue, the converse is actually the case. Subjective collapse, with its retention of unitary dynamics, \emph{does} conserve energy; whereas objective collapse, being necessarily non-unitary, invariably does not.

At the universal level, the expectation energy of the total system (such as is given by \eqref{eq:cons_exptE}) will always be constant. Hence, since under subjective collapse these components are always retained, energy must always be conserved. However, at the relative state level, the energy expectation of the subsystem/relative state product, $\expt{E_{i}}$, is generally different to $\expt{E}$. So, from the point of view of an observer embedded in that relative state, it would seem that energy has \emph{not} been conserved. 

In fact, this is not such a strange conclusion. To the untrained eye, energy always seems to be disappearing, such as when a bouncing ball comes to rest or a cup of coffee cools. As physicists, we reassure the sceptical that this energy has not been lost - it has just been transferred to the environment and is no longer in a usable form. In terms of our current language, we can describe this in terms of losing information about the total system. Here, we will attempt to express this notion quantitatively in terms or relative states (which we shall assume to have become orthogonal).

The energy effectively lost from the point of view of an observer in the $k$th relative state is

\begin{align}
\Delta E_{k} &= \expt{E} - \expt{E_{k}}, \nonumber \\
&= \sum_{i} \sqmod{a_{i}}\expt{E_{i}}  - \expt{E_{k}}. \label{eq:Delta_E_k}
\end{align}

\noindent At this point, we follow the approach of equating the probabilities (given by the squared moduli) to the Boltzmann distribution, putting $\sqmod{a_{i}} = Z^{-1}e^{-\expt{E_{i}}/T}$, where $Z$ is the partition function and $T$ is taken to have dimensions of energy (putting the Boltzmann constant $k_{B} = 1$). We then have $\expt{E_{i}} = -T(\log\sqmod{a_{i}} + \log Z)$. Inserting this into the expression above, we get

\begin{equation}
\frac{\Delta E_{k}}{T} = -\left(\sum_{i} \sqmod{a_{i}}\log\sqmod{a_{i}}  - \log\sqmod{a_{k}}\right). \nonumber
\end{equation}

\noindent On comparison with \eqref{eq:Delta_Sk}, this gives our earlier definition of the change in information with respect to the $k$th relative state and recovers the Clausius form for the thermodynamic entropy

\begin{equation}
\lim_{\iprod{R_{i'}}{R_{i}}\to \delta_{i'i}}\Delta S_{k} = \frac{\Delta E_{k}}{T}. \label{eq:Delta_S_k}
\end{equation}

\noindent (Here, since $\Delta E_{k}$ is effectively dissipated energy as it can no longer be used, we may equate it to the heat $\Delta Q$).

Note that under an objective collapse interpretation, in which all but one relative state disappears along with its energy, the conservation of energy \emph{would be} violated. However, as previously mentioned, objective collapse also breaks the unitary evolution of the Universe and it is out of the time symmetry inherent in this that the conservation of energy emerges in the first place via Noether's theorem~\cite{Noether1918}. 

This conclusion may still sit uncomfortably with those who entirely reject many-worlds-type interpretations. They may still argue that these \emph{do} still violate energy conservation since each possible world would be have an energy of $\expt{E_{i}}$ and that the sum total of these $\sum_{i}\expt{E_{i}}$ will differ from the energy expectation of the total system (as given by \eqref{eq:cons_exptE}). To this, we offer the notion that these energies are not partitioned off individually to each possible world but rather \emph{shared} between them. 

From the point of view of any observer in a relative state, the energy of the others correlates to the thermodynamic entropy of his or her universe. Combining \eqref{eq:Delta_E_k} with \eqref{eq:Delta_S_k}, we may express this in the relation

\begin{equation}
\expt{E_{k}} + T\Delta S_{k} = \expt{E}, \label{eq:total_energy}
\end{equation}

\noindent for all $k$. Here, we may view $\expt{E}$ as the total energy of the combined system, whilst $\expt{E_{k}}$ is analogous to the Helmholtz free energy.

These other `worlds' do not constitute distinctly viable alternatives for the very reason that the remaining energy cannot be partitioned between them. They are, informationally speaking, inconsistent with one another. On the other hand, each still represents a \emph{possible} world. Our argument would then be that a `possible' world is not an `actual' world but rather that collectively they are a superposition of what is \emph{not} the case in one's own world. Philosophically, this might actually be construed as an argument \emph{against} Everett's Many-World Interpretation, although we stress it is \emph{not} an argument against subjective collapse.

%*******************************************************************************************
\section{Discussion}\label{sec:entropic_time}
%*******************************************************************************************
\subsection{The perception of time as information gain}
Using the tools of decoherence theory in conjunction with the relative state formalism, we have seen how information may be transferred from one system to another via entanglement. This has not only given us a deep insight into the connection between entropy and information but presents an explanation of how the classical world of our perceptions arises out of a quantum mechanical description of the Universe. Even more profoundly, we have seen how the apparently non-unitary evolution of a Universe subject to the Second Law of Thermodynamics can emerge from a unitary physics. 

At the same time, we have seen that the actuality of a particular observation corresponds to the transfer of information to a particular relative state. In Section~\ref{sec:info_gain}, we described this information in terms of `entropic time'. Here, the use of the word `time' applies more to our everyday perception of time as a measure of change than it does to the parameter of unitary evolution. 

From the point of view of any cognitive entity embedded in a given relative state, we would argue that both the actuality of an observation and the notion of any change in that state of affairs can \emph{only} be described in terms of information. Consider, for example the relative state of an entity into which a given set of informational facts $X_{0} = \{I_{1}, I_{2},\ldots I_{N}\}$ had been encoded. Any subset of these facts may be taken to be a description of the `past' of that relative state. Suppose that this information was then supplemented by new information $I_{N+1}$. This new set of information $X_{1}$ then represents a \emph{subsequent} state of affairs. Now suppose that somehow the new information $I_{N+1}$ becomes erased. From the point of view of the relative state, we are back in the initial state of affairs, with no information of ever having been otherwise.

In other words, from the point of view of an entity embedded in a relative state, information gain is \emph{effectively} irreversible. It would not matter if, according to the unitary evolution of the system at the universal level, the relative state oscillated between sets $X_{0}$ and $X_{1}$ - from the point of view of the relative state, this would just represent two subsequent states. Such an oscillation could only be perceived by a larger system that repeatedly stored information about the alternating states.

As cognitive entities ourselves, this must be how we actually perceive time, since no such perception would be possible without memory of previous events. It also explains very clearly why we perceive a `now' moment complete with its memories of past events, despite the fact that unitary physics cannot single out a particular point in space or time as being special.

\subsection{Subjective vs objective collapse}\label{sec:subject_object}
\subsubsection{The conservation of information}
Taking quantum mechanics at face value, we have seen that at the universal level, the entropy of the total system is zero and always remains so. This may be taken as a statement of the \emph{conservation of information} at the universal level, since it implies no loss of information from the total system. Such a view must ultimately be sacrificed under any objective collapse interpretation, implying as it does that certain possibilities do become erased somehow. Hence, subjective and objective collapse interpretations do have profoundly different theoretical consequences.

In particular, key differences are:

\begin{itemize}
\item [(1)] Objective collapse requires fundamental new physics beyond our present understanding of quantum mechanics. 
\item [(2)] Information is not conserved under objective collapse.
\item [(3)] Under objective collapse, the evolution of the total system is inherently non-unitary.
\end{itemize}

\subsubsection{No-go theorems}\label{sec:no_go}
On the basis of point (2) above, differences between subjective and objective collapse theories do turn out to have consequences for certain `no-go' theorems of quantum physics\cite{wootters1982single,dewitt2015many,pati2000impossibility}. In particular, the \emph{no deletion} theorem\cite{pati2000impossibility} would appear to be manifestly violated under objective collapse. This theorem is the time-symmetric dual of the \emph{no cloning} theorem\cite{wootters1982single,dewitt2015many}. Both of these theorems follow from the fact that under unitary evolution, whilst information may be transferred from one system to another and manifestly lost due to decoherence, the information is never lost at the level of the total system. The act of actually deleting information would clearly violate this.

A related issue in quantum computing is that of \emph{setting} a qubit of information. In classical computing, setting a bit is a trivial matter - we just overwrite the value of the bit with whatever we want to set it to. However, this is an irreversible operation. Once set, unless we have stored the original value of the bit somewhere else, that information will have been irreversibly deleted. In practice, in order to `set' a qubit, one might allow the two-state system representing it to thermalise with its environment, so that it settled to its lowest energy state. Due to the vast size of the Hilbert space of its environment, this could be made to be an effectively irreversible process. However, under the subjective collapse picture, the information is never actually lost - it just becomes entangled with inaccessible degrees of freedom of the environment.

Under objective collapse, however, the information \emph{would} be lost. Thus, allowing a qubit to thermalise with its environment would represent a \emph{true} deletion process, in violation of the no deletion theorem. This, of course, is only to be expected as the no-go theorems are derived on the basis of unitary evolution and we have acknowledged that objective collapse implies that the temporal evolution of the Universe is inherently non-unitary. It does mean, though, that we can no longer assert these theorems as necessarily true about the Universe. Specifically, since these are \emph{theorems} based on the known postulates of quantum mechanics, if they are falsified, then so are the postulates they are derived from. From this, we may draw the conclusion that whilst subjective collapse theories remain interpretations of quantum mechanics (as it stands), objective collapse theories \emph{are not}, since their truth would falsify the theory they are trying to interpret!

\section{Conclusions}
Since Maxwell's early invocation of his Demon, many great thinkers have devoted themselves to understanding how the non-unitary physics implied by the Second Law of Thermodynamics could possibly arise out time-symmetric dynamical laws. The deep connection between entropy and information hinted at by Maxwell has been long in developing. The ideas developed in this paper are very much influenced by the philosophy that entropy and information are essentially the same thing. In practice, we may differentiate between these by speaking of information as being related to a specific outcome, whilst entropy is the expectation value of the information for different possibilities.

Using the tools of decoherence theory, we have seen how unitary physical laws can lead to the transfer of information between systems via entanglement. Moreover, the precise nature of the interaction between a system and its environment determines which system pointer states are then preferably selected and encoded into the relative environmental state.

The exact fate of relative states containing inconsistent information is still an open question. However, whether the `collapse' of possibilities associated with the gain of specific information is subjective or objective, it is effectively irreversible in either case. This is how we perceive the world around us. Our perception of time is based on the fact that we have information about `past' events. 

We have described this accrual of information in terms of entropic time, which we would argue is far more in keeping with our perceive sense of temporal evolution than the parametric time of unitary physical theory. Indeed, this notion of time is perfectly consistent with the Second Law of Thermodynamics - arguably the most assuredly correct theory in all of physics. 

For practical purposes, subjective and objective collapse interpretations may be similar enough to each other to negate the need for choosing one over the other. Having said that, there are significant consequences for the truth of objective collapse theories. It has been shown that whilst subjective collapse does not violate the conservation of energy, objective collapse does. This is concomitant with the non-unitary nature of objective collapse and its violation of the conservation of information. The implications for this may be more than philosophical. Current efforts to achieve practical quantum computers and other advances in quantum information theory are based on the conservation of information and the validity of the `no-go' theorems. If information is truly lost from the Universe, then then the scalability of such applications is likely to be severely limited.

\appendix
\section{Appendix}
%==========================================================================
\subsection{The Gibbs and Clausius entropies}\label{sec:GibbsClausius}
%==========================================================================
The Gibbs expression for the entropy of a system of $N$ particles is given by \eqref{eq:Gibbs} in terms of the probabilities $p_{i}$ for a particular micro-state. In the following, we shall assume the number of particles $N$ remains constant. We further assume that the probabilities are given by Boltzmann factor \eqref{eq:Boltz_factor}. The total energy $E_{i}$ is the sum of kinetic and potential energy components, which we shall denote by $K_{i}$ and $U_{i}$ respectively, i.e. $E_{i} = K_{i} + U_{i}$.

The expectation energy of the system is given by

\begin{equation}
\expt{E} = \sum_{i}p_{i}E_{i},  \label{eq:E_expt}
\end{equation}

\noindent so an infinitesimal change in this will be given by

\begin{equation}
d\expt{E} = \sum_{i}d(p_{i}E_{i}) = \sum_{i}\left(dp_{i}E_{i} + p_{i}dE_{i}\right).  \label{eq:dE_expt}
\end{equation}

\noindent Similarly, a small change in $S$ is given by

\begin{equation}
dS = -k_{B}\sum_{i}^{\Omega}\left(dp_{i}\log p_{i} + dp_{i}\right) = -k_{B}\sum_{i}^{\Omega}dp_{i}\log p_{i}, \nonumber
\end{equation}

\noindent since, by the conservation of probability, $\sum_{i}dp_{i} = 0$. Using $p_{i} = Z^{-1}e^{-E_{i}/(k_{B}T)}$ from \eqref{eq:Boltz_factor}, this becomes

\begin{equation}
dS = \frac{1}{T}\sum_{i}^{\Omega}dp_{i}E_{i}, \nonumber
\end{equation}

\noindent where the term involving $\log Z$ has disappeared, again due to the conservation of probability. 

Substituting from \eqref{eq:dE_expt} then gives

\begin{equation}
dS = \frac{1}{T}\left(d\expt{E} - \sum_{i}p_{i}dE_{i}\right). \label{eq:dS1}
\end{equation}

Now

\begin{equation}
dE_{i} = dK_{i} + dU_{i} = dK_{i} - P_{i}dV, \nonumber
\end{equation}

\noindent where $P_{i}$ is the pressure of the $i$th state and $dV$ is the change in volume. $K_{i}$ will be proportional to the number of particles in the system $N$ and the temperature, so 

\begin{equation}
dK_{i} \propto TdN + NdT = NdT, \nonumber
\end{equation}

\noindent since, by assumption, the number of particles does not change. 

We shall now make use of the path independence of macroscopic thermodynamic states and follow a constant-temperature, reversible path from initial to final state. In this case, we have $dT = 0$, so in the sum of $p_{i}dE_{i}$ over all configurations, we are left with

\begin{equation}
\sum_{i}p_{i}dE_{i} = -\sum_{i}p_{i}P_{i}dV = \expt{dW}, \nonumber
\end{equation}

\noindent i.e. the expectation value of the work done on the system.

Substituting this into \eqref{eq:dS1}, we have

\begin{equation}
dS = \frac{d\expt{E} - \expt{dW}}{T}.
\end{equation}

\noindent However, from the first law of thermodynamics, we have $d\expt{E} = \expt{dQ} + \expt{dW}$, where $\expt{dQ}$ is the expectation value of the heat transfer, so we arrive at

\begin{equation}
dS = \frac{\expt{dQ}}{T}, \nonumber
\end{equation}

\noindent which is in the form of Clausius' expression \eqref{eq:Clausius} for the phenomenological thermodynamic entropy.

%==========================================================================
\subsection{Properties of the density matrix}
%==========================================================================
\subsubsection{Positive semi-definiteness of the density matrix}\label{sec:positive_semi_def}
We show here that any density matrix is positive semi-definite - that is, that its eigenvalues are always greater or equal to zero.

\paragraph{Pure state density matrix:\\}

For a pure state density matrix, we have

\begin{equation}
\rho = \ket{\Psi}\bra{\Psi}. \label{eq:pure_state_rho}
\end{equation}

\noindent Suppose that $\ket{\theta_{i}}$ is an eigenvector of $\rho$ with eigenvalue $\lambda_{i}$, so

\begin{equation}
\rho\ket{\theta_{i}} = \lambda_{i}\ket{\theta_{i}}. \nonumber
\end{equation}

\noindent Taking the inner product of this with $\bra{\theta_{i}}$, we have

\begin{equation}
\iprodM{\theta_{i}}{\rho}{\theta_{i}} = \lambda_{i}. \label{eq:iprod_pure_state}
\end{equation}

\noindent Substituting for $\rho$ from \eqref{eq:pure_state_rho}, we then find

\begin{equation}
\lambda_{i} = \iprod{\theta_{i}}{\Psi}\iprod{\Psi}{\theta_{i}} = \sqmod{\iprod{\theta_{i}}{\Psi}} \ge 0. \nonumber
\end{equation}

\noindent Hence, the eigenvalues of a pure state density matrix are always greater or equal to zero.

\paragraph{Mixed state density matrix:\\}

For a mixed state density matrix, we have a matrix of the form

\begin{equation}
\rho = \sum_{k}p_{k}\ket{\Psi_{k}}\bra{\Psi_{k}}, \label{eq:mixed_state_rho}
\end{equation}

\noindent where the $p_{i}$ are probabilities and so greater or equal to zero. Inserting this into \eqref{eq:iprod_pure_state}, we have

\begin{equation}
\lambda_{i} = \sum_{k}p_{k}\iprod{\theta_{i}}{\Psi_{k}}\iprod{\Psi_{k}}{\theta_{i}} = \sum_{k}p_{k}\sqmod{\iprod{\theta_{i}}{\Psi_{k}}} \ge 0, \nonumber
\end{equation}

\noindent as before.

\paragraph{Reduced density matrix:\\}

For a density matrix

\begin{equation}
\rho = \sum_{ii'jj'} \ket{\f_{i}\E_{j}}a_{ij}a_{i'j'}^{*}\bra{\E_{j'}\f_{i'}}, \nonumber
\end{equation}

\noindent we may form the reduced density matrix

\begin{equation}
\rho_{R} = \sum_{j}\iprodM{\E_{j}}{\rho}{\E_{j}} = \sum_{j}\sum_{ii'} \ket{\f_{i}}a_{ij}a_{i'j}^{*}\bra{\f_{i'}}. \nonumber
\end{equation}

As before, for an eigenvector of $\rho_{R}$, we have

\begin{equation}
\lambda_{k} = \iprodM{\theta_{k}}{\rho_{R}}{\theta_{k}}  = \sum_{j}\sum_{ii'} \iprod{\theta_{k}}{\f_{i}}a_{ij}a_{i'j}^{*}\iprod{\f_{i'}}{\theta_{k}}. \nonumber
\end{equation}

\noindent Putting $b_{ki} = \iprod{\theta_{k}}{\f_{i}}$, this is

\begin{equation}
\lambda_{k} = \sum_{j}\sum_{ii'} b_{ki}a_{ij}a_{i'j}^{*}b_{ki'}^{*}. \nonumber
\end{equation}

Defining a new vector

\begin{equation}
\ket{\Xi_{k}} = \sum_{j}A_{kj}\ket{\xi_{j}}, \nonumber
\end{equation}

\noindent where

\begin{equation}
A_{kj} = \sum_{i}b_{ki}a_{ij}, \nonumber
\end{equation}

\noindent we see that 

\begin{equation}
\iprod{\Xi_{k}}{\Xi_{k}} = \sum_{j}\sqmod{A_{kj}} = \sum_{j}\sum_{ii'} b_{ki}a_{ij}a_{i'j}^{*}b_{ki'}^{*} = \lambda_{k} \ge 0.  \nonumber
\end{equation}

\subsubsection{Zero entropy of a pure state density matrix}\label{sec:zero_entropy}
In Section~\ref{sec:positive_semi_def} above, it was shown that the density matrix is positive semi-definite, meaning that its eigenvalues are always greater or equal to zero. 

Furthermore, it is evident, from the definition of a density matrix, that the trace is always equal to unity. For a pure state matrix formed from the state

\begin{equation}
\ket{\Psi} = \sum_{i}a_{i}\ket{\phi_{i}}, \nonumber
\end{equation}

\noindent the trace of $\rho =\ket{\Psi}\bra{\Psi}$ is

\begin{equation}
\mathrm{Tr}\left[\rho\right] = \sum_{i}\sqmod{a_{i}} = 1, \nonumber
\end{equation}

\noindent due to the normalisation of $\ket{\Psi}$, whilst for a mixed-state density matrix such as \eqref{eq:mixed_state_rho}, we have

\begin{equation}
\mathrm{Tr}\left[\rho\right] = \sum_{k}p_{k} = 1, \nonumber
\end{equation}

\noindent since the probabilities must sum to unity.

For a pure-state density matrix $\rho =\ket{\Psi}\bra{\Psi}$, it is straight-forward to show that $\ket{\Psi}$ itself is an eigenvector of $\rho$ with eigenvalue unity:

\begin{equation}
\rho\ket{\Psi} = \ket{\Psi}\iprod{\Psi}{\Psi} = \ket{\Psi}. \nonumber
\end{equation}

\noindent Now the trace of any square, invertible matrix is invariant and equal to the sum of its eigenvalues. For a density matrix this is unity. Since the density matrix is also positive semi-definite, this means that all other eigenvalues must be equal to zero. 

The von Neumann entropy will then be a sum of $0 \log 0 \equiv 0$ terms plus a single term $\log 1 = 0$. That is,

\begin{equation}
S\left(\ket{\Psi}\bra{\Psi}\right) = \sum_{i} \lambda_{i}\log\lambda_{i} = 0. \nonumber
\end{equation}

\noindent Thus, the entropy of pure density matrix is identically zero.

%==========================================================================
\subsection{Entropy of entanglement in a bipartite system}\label{app:entropy_entanglement}
%==========================================================================
In sections~\ref{sec:bipartite_open} and \ref{sec:reduced_density}, we considered a bipartite system consisting of a system and its environment. Here, we adopt a more general description of a bipartite system considering of two subsystems $\S_{1}$ and $\S_{2}$. A general state of this system may be written in the form of \eqref{eq:Psi_rel} where the $\ket{\f_{i}}$ are taken to be eigenvectors of $\rho_{1}$, the reduced density matrix of $\S_{1}$. This is then in the Schmidt decomposition with $\iprod{\R_{i}}{\R_{j}} = \delta_{ij}$.

The entropy of $\S_{1}$ may then be written down as

\begin{equation}
S(\rho_{1}) = - \sum_{k}\sqmod{a_{k}}\log\sqmod{a_{k}}. \label{eq:S_rho1}
\end{equation}

Rewriting the total density matrix in terms of the relative states (which are now mutually orthogonal) using \eqref{eq:Psi_rel}, we have

\begin{equation}
\rho_{12} = \sum_{i'i}\ket{\f_{i}\R_{i}}a_{i}a_{i'}^{*}\bra{\f_{i}\R_{i}}. \nonumber
\end{equation}

\noindent Using this to calculate the reduced density matrix of $\S_{2}$, we find

\begin{align}
\rho_{2} &= \sum_{k}\iprodM{\f_{k}}{\rho_{12}}{\f_{k}}, \nonumber \\
&= \sum_{k}\ket{\R_{k}}\sqmod{a_{k}}\bra{\R_{k}}. \nonumber
\end{align}

\noindent Hence, we see that, in this case, the $\ket{\R_{k}}$ are eigenvectors of $\rho_{2}$ with eigenvalues $\sqmod{a_{k}}$ and the entropy of $\S_{2}$ is

\begin{equation}
S(\rho_{2}) = - \sum_{k}\sqmod{a_{k}}\log\sqmod{a_{k}} = S(\rho_{1}). \label{eq:bipartite_theorem}
\end{equation}

\noindent Since the entropy of the subsystems does not depend of the basis chosen to represent them, \eqref{eq:bipartite_theorem} is a theorem for any bipartite system. 

%==========================================================================
\subsection{Time derivative of the reduced density matrix}\label{app:drho_dt}
%==========================================================================
The temporal evolution of the total state $\ket{\Psi}$ is found from

\begin{equation}
\frac{d}{dt}\ket{\Psi} = -\frac{\rmi}{\hbar}H\ket{\Psi}. \label{eq:temporal_evolve}
\end{equation}

\noindent Representing $\ket{\Psi}$ in terms of relative states, for a bipartite system with $H$ given by \eqref{eq:H_form}, we have

\begin{align}
\frac{d}{dt}\ket{\Psi} &= \sum_{k}\left(\frac{da_{k}}{dt}\ket{\f_{k}\R_{k}} + a_{k}\ket{\f_{k}}\frac{d}{dt}\ket{\R_{k}}\right), \nonumber \\
&= -\frac{\rmi}{\hbar}\sum_{k}a_{k}\left(H_{\S} + H_{\Env} + H_{\S\Env}\right)\ket{\f_{k}\R_{k}}. \label{eq:temp2}
\end{align}

We shall choose system states $\ket{\f_{i}}$ and environmental states $\ket{\E_{j}}$ such that  $H_{\S\Env}$ is diagonal in $\ket{\f_{i}\E_{j}}$. We may then put

\begin{equation}
\iprodM{\f_{i}}{H_{\S\Env}}{\f_{k}\R_{k}} = \delta_{ik}H_{\S\Env}^{i}\ket{\R_{i}}, \nonumber
\end{equation}

\noindent where $H_{\S\Env}^{i} \equiv \iprodM{\f_{i}}{H_{\S\Env}}{\f_{i}}$.

Taking the inner product of \eqref{eq:temp2} with $\bra{\f_{i}}$ then gives

\begin{align}
\frac{da_{i}}{dt}\ket{\R_{i}} + a_{i}\frac{d}{dt}\ket{\R_{i}} &= -\frac{\rmi}{\hbar}\sum_{k}a_{k}\iprodM{\f_{i}}{H_{\S}}{\f_{k}}\ket{\R_{k}} \nonumber \\
&-\frac{\rmi}{\hbar}a_{i}\left(H_{\Env} + H_{\S\Env}^{i}\right)\ket{\R_{i}}. \nonumber  
\end{align}

\noindent Dividing through by $a_{i}$ and rearranging

\begin{align}
 & \frac{1}{a_{i}}\frac{da_{i}}{dt}\ket{\R_{i}} + \frac{\rmi}{\hbar}\sum_{k}\frac{a_{k}}{a_{i}}\iprodM{\f_{i}}{H_{\S}}{\f_{k}}\ket{\R_{k}}  \nonumber \\
=  &-\left[\frac{\rmi}{\hbar}\left(H_{\Env} + H_{\S\Env}^{i}\right)\ket{\R_{i}} + \frac{d}{dt}\ket{\R_{i}}\right]. \nonumber  
\end{align}

The right-hand-side of this is now independent of the coefficients $a_{i}$, so for this equation to hold generally, both sides must be identically zero. Hence, we have

\begin{align}
\frac{da_{i}}{dt}\ket{\R_{i}} = -\frac{\rmi}{\hbar}\sum_{k}a_{k}\iprodM{\f_{i}}{H_{\S}}{\f_{k}}\ket{\R_{k}}. \nonumber  
\end{align}

\noindent Multiplying on the left by $a_{j}^{*}\bra{\R_{j}}$ and putting $H_{ij} = \iprodM{\f_{i}}{H_{\S}}{\f_{j}}$,

\begin{align}
\frac{da_{i}}{dt}a_{j}^{*}\iprod{\R_{j}}{\R_{i}} = -\frac{\rmi}{\hbar}\sum_{k}H_{ik}a_{k}a_{j}^{*}\iprod{\R_{j}}{\R_{k}}. \nonumber  
\end{align}

\noindent Here, we recognise the elements of the reduced density matrix $\rho_{\S}$ for the subsystem

\begin{equation}
\rho_{ij} \equiv \iprodM{\f_{i}}{\rho_{\S}}{\f_{j}} = a_{i}a_{j}^{*}\iprod{\R_{j}}{\R_{i}}, \label{eq:rho_ij}
\end{equation}

\noindent so we have

\begin{align}
\frac{da_{i}}{dt}a_{j}^{*}\iprod{\R_{j}}{\R_{i}} &= -\frac{\rmi}{\hbar}\sum_{k}H_{ik}\rho_{kj}, \nonumber \\
& = -\frac{\rmi}{\hbar}\iprodM{\f_{i}}{H_{\S}\rho_{\S}}{\f_{j}}. \label{eq:H_rho}
\end{align}

Since both $H_{\S}$ and $\rho_{\S}$ are Hermitian, taking the complex conjugate of this gives

\begin{align}
\frac{da_{i}^{*}}{dt}a_{j}\iprod{\R_{i}}{\R_{j}} &=\frac{\rmi}{\hbar}\sum_{k}\rho_{jk}H_{ki}, \nonumber \\
& = \frac{\rmi}{\hbar}\iprodM{\f_{j}}{\rho_{\S}H_{\S}}{\f_{i}}, \nonumber
\end{align}

\noindent or, on swapping indices,

\begin{align}
a_{i}\frac{da_{j}^{*}}{dt}\iprod{\R_{j}}{\R_{i}} &= \frac{\rmi}{\hbar}\iprodM{\f_{i}}{\rho_{\S}H_{\S}}{\f_{j}}. \label{eq:rho_H} 
\end{align}

Taking the time derivative of \eqref{eq:rho_ij} and using \eqref{eq:H_rho} and \eqref{eq:rho_H}, we then find

\begin{align}
\frac{d\rho_{ij}}{dt} &= \left(\frac{da_{i}}{dt}a_{j}^{*} + a_{i}\frac{da_{j}^{*}}{dt}\right)\iprod{\R_{j}}{\R_{i}} +  a_{i}a_{j}^{*}\frac{d}{dt}\iprod{\R_{j}}{\R_{i}}, \nonumber \\
&= -\frac{\rmi}{\hbar}\iprodM{\f_{i}}{\left[H_{\S},\rho_{\S}\right]}{\f_{j}} +  \frac{\rho_{ij}}{\iprod{\R_{j}}{\R_{i}}}\frac{d}{dt}\iprod{\R_{j}}{\R_{i}}. \nonumber
\end{align}

\noindent Recognising the logarithmic derivative, this may be written

\begin{align}
\frac{d\rho_{ij}}{dt} &= -\frac{\rmi}{\hbar}\iprodM{\f_{i}}{\left[H_{\S},\rho_{\S}\right]}{\f_{j}} +  \rho_{ij}\frac{d}{dt}\log\iprod{\R_{j}}{\R_{i}}. \nonumber
\end{align}

%==========================================================================
\subsection{Relative state inner products in the diagonal energy model}
%==========================================================================
\subsubsection{Step function}\label{app:sinc}
We saw in section~\ref{sec:DEM_ortho} that the inner product of two distinct relative states in the diagonal energy model is given by \eqref{eq:R_inner}. Let us assume that $\w_{i'j} - \w_{ij}$ varies linearly with $j$ and that the energy differences between terms in the summation over $j$ are small enough that we may move to the continuum limit

\begin{equation}
\iprod{\R_{i'}}{\R_{i}} = \sum_{j}\sqmod{b_{j}}\rme^{\rmi(\w_{i'j} - \w_{ij})t} \to \int \sqmod{b_{\w}}\rme^{\rmi\w t}d\w, \label{eq:inner_prod_app}
\end{equation}

\noindent where we have rescaled $b$ to give it dimensions of time and used $\w = \w_{i'j} - \w_{ij}$ for brevity.

Now, let us further assume that $b_{\w}$ is constant over a bandwidth $\Omega$ centred on $\Omega_{0}$ and zero outside of it. At $t=0$, the normalisation of the eigenvectors means that

\begin{equation}
\int_{\Omega_{0}-\Omega/2}^{\Omega_{0} + \Omega/2} \sqmod{b_{\w}}d\w = \sqmod{b_{\w}}\Omega = 1, \nonumber
\end{equation}

\noindent so $\sqmod{b_{\w}} = 1/\Omega$. For $t > 0$, we then have

\begin{equation}
\frac{1}{\Omega}\int_{\Omega_{0}-\Omega/2}^{\Omega_{0}+\Omega/2} \rme^{\rmi\w t}d\w = \frac{2\rme^{i\Omega_{0}t}}{\Omega t}\sin\left(\frac{\Omega t}{2}\right) = \rme^{i\Omega_{0}t}\sinc\left(\frac{\Omega t}{2}\right). \nonumber
\end{equation}

\noindent Hence the inner product becomes zero after a characteristic time period $\tau = 2\pi/\Omega$ with decreasing amplitude oscillations thereafter.

\subsubsection{Thermal distribution}\label{app:thermal}
An alternative model for the environmental states is to assume a thermal distribution $\sqmod{b_{\w}} = \tau\rme^{-\w\tau}$, where $\tau = \hbar/(k_{B}T)$ in terms of the thermal energy $k_{B}T$. Assuming that $\w_{i'j} - \w_{ij} \propto \w$ and absorbing any constant of proportionality into the time units, \eqref{eq:inner_prod_app} for the relative state inner product becomes

\begin{align}
\iprod{\R_{i'}}{\R_{i}} &= \tau\int_{0}^{\infty} \rme^{\left(\rmi t - \tau\right)\w}d\w, \nonumber \\
&= \frac{1}{1 - \rmi t/\tau}, \nonumber \\
&= \frac{\rme^{\rmi\theta(t)}}{\left(1 + t^{2}/\tau^{2}\right)^{1/2}}, \nonumber 
\end{align}

\noindent where $\theta(t) = \tan^{-1}(t/\tau)$.

\bibliographystyle{unsrt}
\bibliography{entropic}

\end{document}